\pdfoutput=1
\documentclass{JINST}

\usepackage{graphicx}
\usepackage{subfigure}

\title{Characterization of new hybrid pixel module concepts for the \mbox{ATLAS} Insertable B-Layer upgrade}

\author{Malte Backhaus$^a$ on behalf of the \mbox{ATLAS} \mbox{IBL} collaboration\\
\llap{$^a$}Physikalisches Institut der Unversit\"{a}t Bonn,\\
  Nu{\ss}allee 12, 53115 Bonn, Germany\\
  E-mail: \email{backhaus@physik.uni-bonn.de}}

\abstract{The \mbox{ATLAS} Insertable B-Layer (\mbox{IBL}) collaboration plans to insert a fourth pixel layer inside the present Pixel Detector to recover from eventual failures in the current pixel system, especially the b-layer. Additionally the \mbox{IBL} will ensure excellent tracking, vertexing and b-tagging performance during the LHC phase I and add robustness in tracking with high luminosity pile-up. The expected peak luminosity for \mbox{IBL} is $2$ to $3 \cdot 10^{34}$ cm$^{-2}$s$^{-1}$ and \mbox{IBL} is designed for an integrated luminosity of 700 fb$^{-1}$. This corresponds to an expected fluence of $5 \cdot 10^{15}$ 1 MeV n$_{eq}$cm$^{-2}$ and a total ionizing dose of 250 MRad.

In order to cope with these requirements, two new module concepts are under investigation, both based on a new front end IC, called FE-I4. This IC was designed as readout chip for future \mbox{ATLAS} Pixel Detectors and its first application will be the \mbox{IBL}. The planar pixel sensor (\mbox{PPS}) based module concept benefits from its well understood design, which is kept as similar as possible to the design of the current \mbox{ATLAS} Pixel Detector sensor. The second approach of the new three dimensional (\mbox{3D}) silicon sensor technology benefits from the shorter charge carrier drift distance to the electrodes, which completely penetrate the sensor bulk. Prototype modules of both sensor concepts have been build and tested in laboratory and test beam environment before and after irradiation. Both concepts show very high performance even after irradiation to $5 \cdot 10^{15}$ 1 MeV n$_{eq}$cm$^{-2}$ and meet the IBL specifications in terms of hit efficiency being larger than 97\%. Lowest operational threshold studies have been effected and prove independent of the used sensor concept the excellent performance of FE-I4 based module concepts in terms of noise hit occupancy at low thresholds.}


\begin{document}

\section{The FE-I4 IC Architecture}
\label{sec:fei4}

Future \mbox{ATLAS} \cite{atlas} Pixel Detector upgrades will need to withstand high radiation doses and need to be capable of coping with very high hit occupancies at increased LHC luminosity, in particular for the foreseen new b-layer radius of ~3.3 cm away from the beam. The current Pixel Detector \cite{aad} readout chip is not capable of coping with the expected hit occupancies and radiation doses \cite{arutinov}, so a completely new readout chip architecture was designed \cite{barbero}. This IC is built in a 130 nm CMOS feature size using thin gate oxide transistors to increase radiation hardness. The large IC ($20.2$ x $18.8$ mm$^2$) has an active area holding 80 columns with 336 pixels each and a ~2 mm high periphery, which results in an active over inactive area fraction of about 90 \%, leading to simplified module concepts.
\begin{figure}[htp]
\centering
  \includegraphics[width=0.8\linewidth]{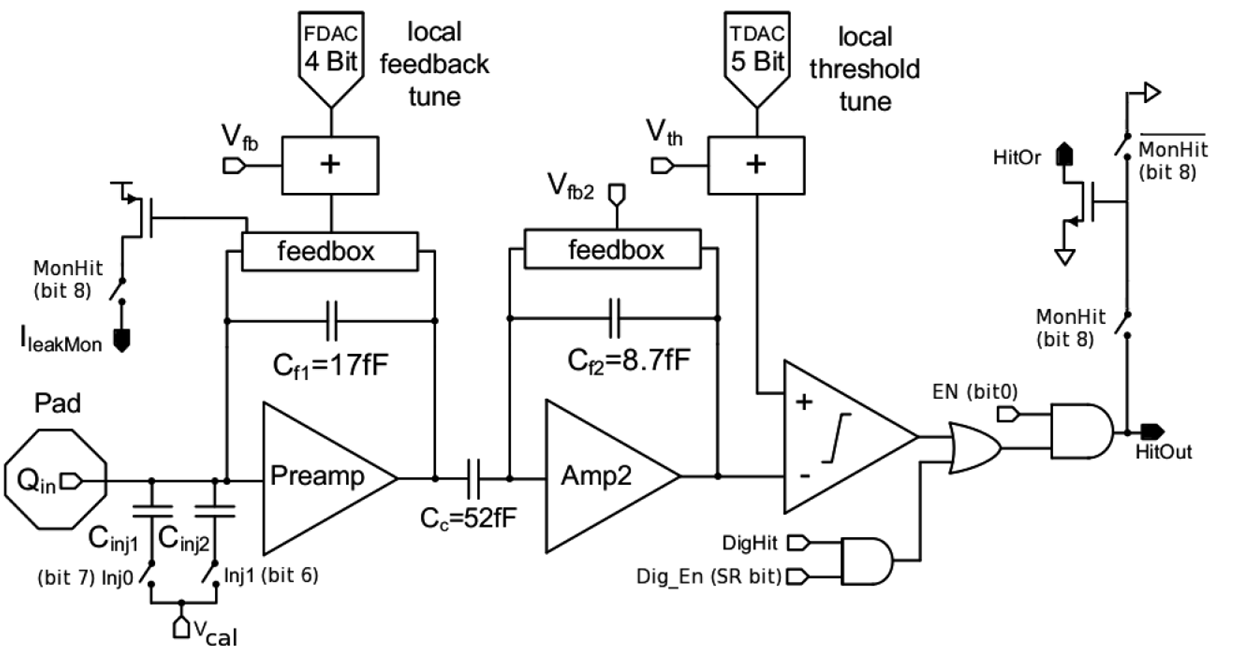}
  \caption{Schematic view of the analog pixel cell.}
  \label{fig:analog}
\end{figure}
All pixels have a size of $250$ x $50$ $\mu$m$^2$ holding a two stage amplifier analog pixel design shown in figure \ref{fig:analog} and four pixels share a common digital logic cell which is  the focus of section \ref{subsec:digitalregion}.

\subsection{The 4-pixel digital region}
\label{subsec:digitalregion}

\begin{figure}[htp]
\centering
  \includegraphics[width=0.8\linewidth]{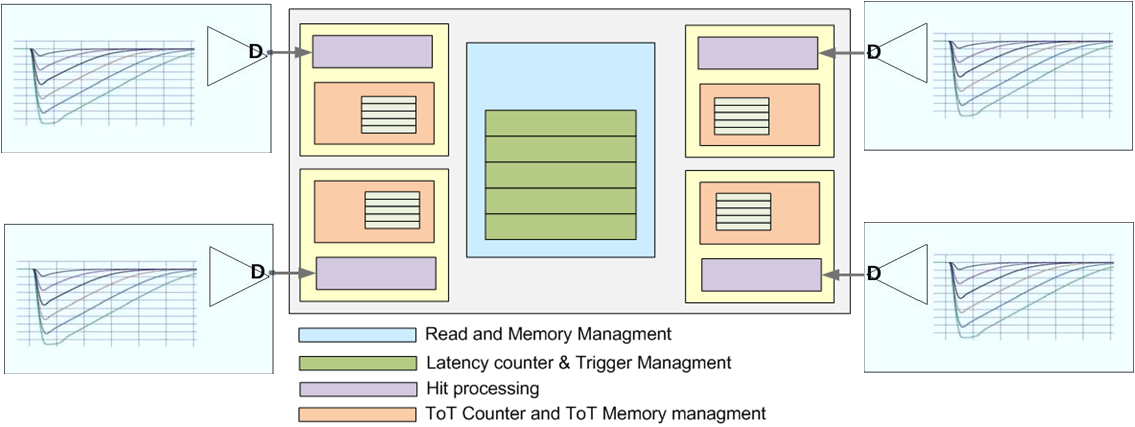}
  \caption{Schematic view of the digital 4-pixel region.}
  \label{fig:4pdr}
\end{figure}

The new digital hit processing is based on the 4-pixel digital region. Detailed studies showed that the transfer of hit information to the chip periphery was the main inefficiency source at the expected \mbox{IBL} hit occupancy \cite{arutinov}. The new hit processing architecture therefore stores the hits in the pixel array close to the analog readout chain. Detailed information on this architecture can be found in \cite{barbero2}. The 4-pixel digital region is sketched in figure \ref{fig:4pdr}.
Four analog pixels share a set of five latency counters, each pixel holding his own set of five Time over Threshold (ToT) counters. A hit in one of the analog pixel cells allocates and starts the first unallocated latency counter to count down from latency (in units of 25 ns). The charge information according to this specific time stamp is stored in the buffers for all pixels connected to the 4-pixel digital region. An arriving level 1 trigger in coincidence with the latency value of zero initiates the transfer of the hit data to the end of chip logic and deallocates the latency counters and buffers. If no corresponding LVL1 trigger arrives, the hit information is deleted and the counter and buffers are deallocated as well. This logic mirrors the clustered nature of real hits.

\subsection{Specifics of the FE-I4A prototype chip and changes for FE-I4B}
\label{subsec:fe-i4a}

FE-I4A is the first full scale prototype chip using this digital readout logic architecture. In this prototype chip, several flavors of the pixel cells have been tested:
\begin{enumerate}
  \item other feedback capacitors in ten out of the forty double columns.
  \item low power discriminators in one double column.
  \item eight double columns with single event upset hard pixel memory cells.
  \item low power discriminators in one double column.
  \item eight double columns with single event upset hard pixel memory cells.
\end{enumerate}
No injection capacitance measurement circuitry was implemented in FE-I4A, which has consequences on the knowledge of the absolute chip calibration. Note that section \ref{sec:calibration} introduces a successful method for measuring the chip calibration constants.\\
The next version of the IC, named FE-I4B, is the \mbox{IBL} production chip, and was submitted for production in September 2011. Of particular importance for this paper, FE-I4B will now hold an uniform pixel array and an injection capacitance measurement circuitry was implemented.

\section{Basic IC and module performance}
\label{sec:basicIC}

In depth IC characterization started directly after the chip returned from production on wafer as well as on diced bare ICs. First results demonstrating the performance of FE-I4 based module prototypes will be shown here. All results were obtained using the \mbox{USBpix} test system \cite{backhaus}.

\subsection{Basic DAC characterizations}
\label{subsec:basicdacs}

\begin{figure}[htp]
\centering
    \includegraphics[width=0.8\linewidth]{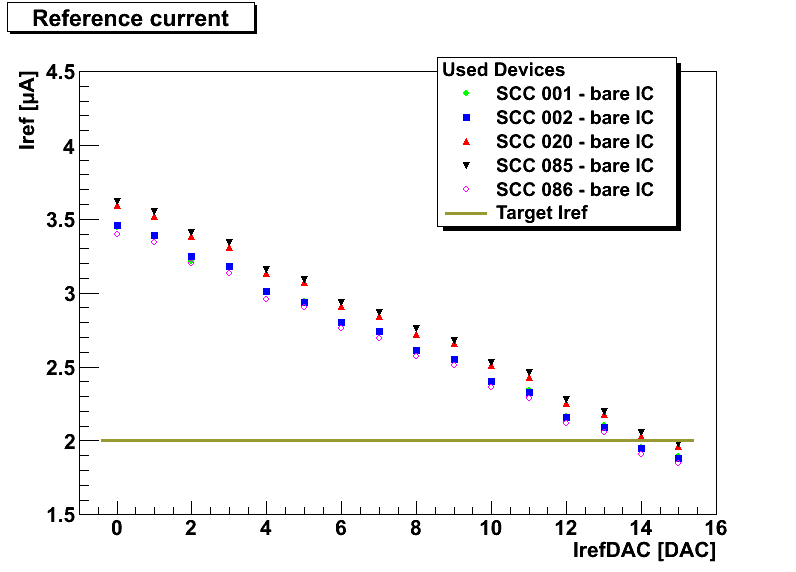}
    \caption{The reference current DAC response of FE-I4A.}
    \label{fig:irefchar}
\end{figure}

The FE-I4 IC generates a current used as reference for all digital to analog converters (DAC). This reference current can be measured and needs to be adjusted to the design value of 2 $\mu$A using a dedicated register in the global memory of the IC. The reference current setting was characterized on several bare ICs and the result is plotted in figure \ref{fig:irefchar}.
For all tested ICs the design reference current of 2 $\mu$A is within the dynamic DAC range, but is on the edge of the trimming range. The DAC range will be readjusted in FE-I4B to achieve a better centering of this range.\\
\begin{figure}[htp]
\centering
    \includegraphics[width=0.8\linewidth]{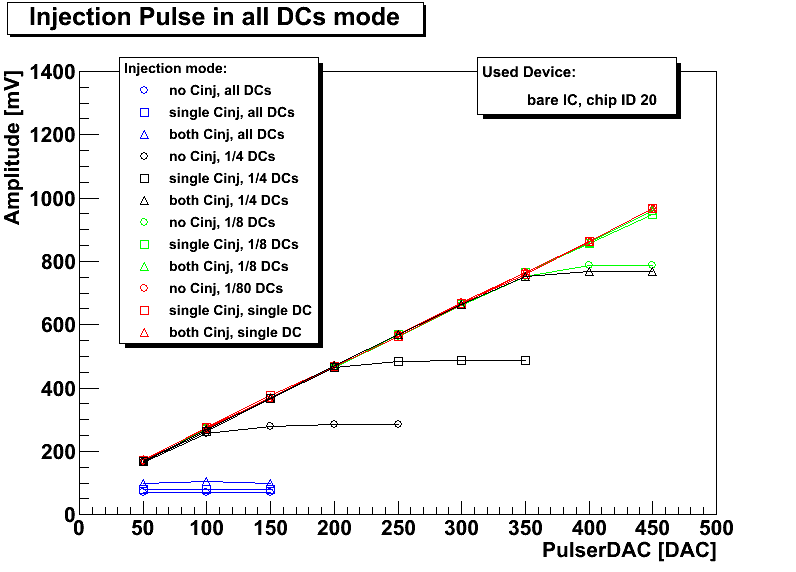}
    \caption{The test injection pulser amplitude for all injection modes.}
    \label{fig:pulserdacchar}
\end{figure}
A test pulse injection block is implemented in the periphery and allows to study the response of the analog readout chain to charge pulses at the input of the preamplifier. The pulser DAC defines a voltage step on one/two injection capacitors implemented in every pixel. A method measuring the injection capacitance itself is introduced in section \ref{sec:calibration}. The voltage jump amplitudes for all possible injection modes as a function of the pulser DAC setting is plotted in figure \ref{fig:pulserdacchar}. Clearly visible is a saturation for high pulser DAC values. The more double columns are selected for injection at the same time the smaller the range available before the pulser DAC saturates. Note also that the saturation occurs at a smaller amplitude when the user selects one capacitor for test pulse injection instead of two. No dependence on the capacitance selected was observed (not shown in this plot). The reason for this behavior is the following: the switches selecting the injection capacitors have a high leakage current to ground when opened. Also the impedance of the pulser block is too high. Despite the described behavior, the use of the one out of eight double column mode together with the use of both injection capacitors is possible, which was consequently implemented as the default scanning procedure in the \mbox{USBpix} test system. This constraint decreases the speed of all measurements using the test pulse injection logic in FE-I4A, for example to perform threshold scans. In the FE-I4B this issue is addressed by using low leakage current switches and lowering the output impedance of the pulser. Higher scan performances is therefore expected in FE-I4B.

\subsection{Prototype module performance}
\label{subsec:moduleperformance}

Prototype modules using planar sensors \cite{wittig} and \mbox{3D} sensors from the two manufactures CNM \cite{pellegrini} and FBK \cite{dalla} have been built, tested and irradiated to the IBL design fluence of $5 \cdot 10^{15}$ 1 MeV n$_{eq}$cm$^{-2}$ using 26 MeV protons in Karlsruhe and using thermal neutrons in Ljubljana. The samples have been unpowered and have not been cooled during the irradiation periods, as the target fluence was reached within a few minutes in both cases. Selected results obtained with these prototype modules are shown here and in section \ref{sec:calibration} and \ref{sec:lowthresholds}. Several unirradiated and irradiated FE-I4A based prototype modules using both sensor technologies have been successfully operated at thresholds down to 1600 electrons in the IBL June test beam environment. More information about this irradiation campaigns, test beam environments and results can be found in \cite{weingarten}.
\begin{figure}[htp]
\centering
    \subfigure[]{
        \includegraphics[width=0.45\linewidth]{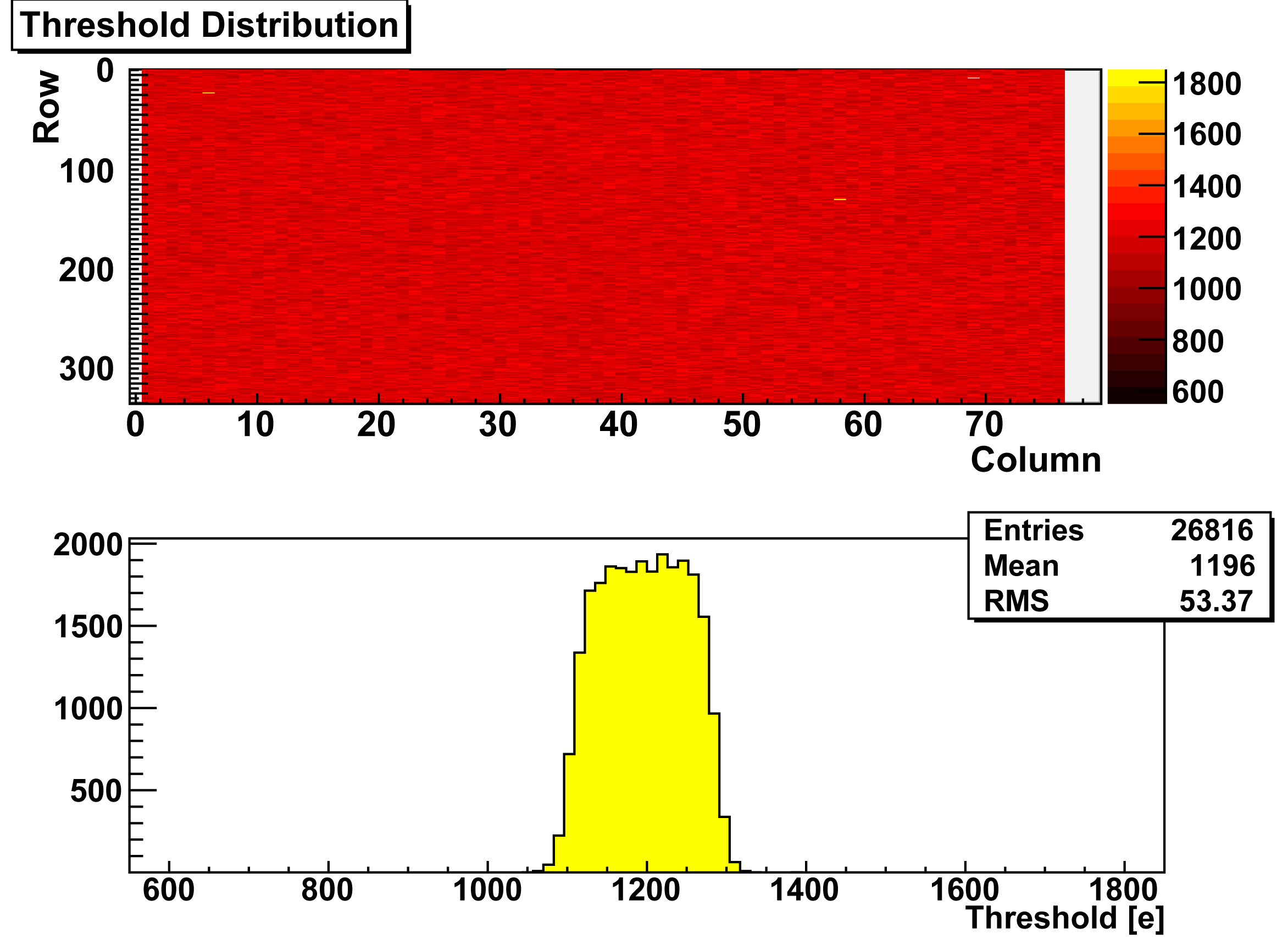}
        \label{fig:threshold_dist}
    }
    \subfigure[]{
        \includegraphics[width=0.45\linewidth]{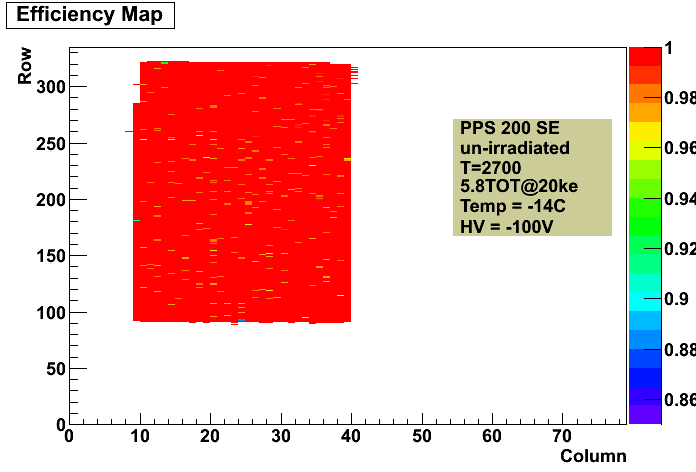}
        \label{fig:hiteff}
    }
    \label{fig:basicperform}
    \caption{Performance of unirradiated planar prototype modules in test beam and lab environment. (a) shows the threshold distribution for a module tuned to 1200 electrons threshold and (b) shows a hit efficiency map obtained at \mbox{IBL} June test beam using 180 GeV pions.}
\end{figure}

An example threshold distribution of a planar pixel prototype module tuned to 1200 electrons threshold is shown in figure \ref{fig:threshold_dist}. The threshold distribution is very narrow and the module can be operated at this low threshold range (more details are provided in section \ref{sec:lowthresholds}).
A hit efficiency map of an unirradiated planar module obtained with 180 GeV pions in \mbox{IBL} June test beam at CERN\footnote{Conseil Européen pour la Recherche Nucléaire} demonstrates the excellent hit efficiency of FE-I4A based prototype modules.

\section{Calibration of the FE-I4}
\label{sec:calibration}

In the FE-I4A IC, no circuitry for direct measurement of the test charge injection capacitances has been implemented. Simulations predict a capacitance of 5.7 fF if both injection capacitors are used. In order to achieve a good absolute calibration of the chip, the pulser DAC calibration (see chapter \ref{subsec:basicdacs}) as well as the injection capacitance need to be known. A method allowing to measure the injection capacitance without using the ToT information (which calibration is tuned using the injection capacitance) and the results of the first example realization on a \mbox{PPS} assembly is shown here.

\subsection{Calibration constant measurement procedure}
\label{sub:calibmeasproc}

\begin{figure}[htp]
\centering
    \subfigure[]{
        \includegraphics[width=0.8\linewidth]{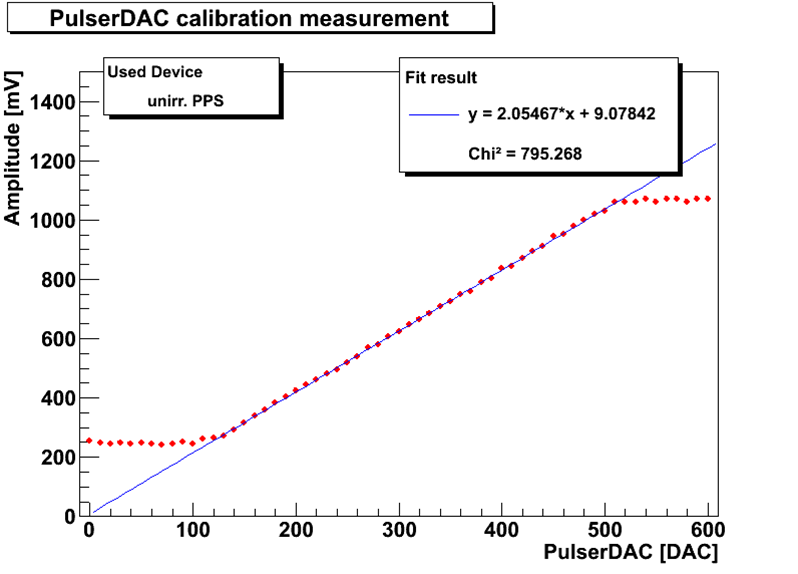}
        \label{fig:pulsercalib}
    }
    \subfigure[]{
        \includegraphics[width=0.8\linewidth]{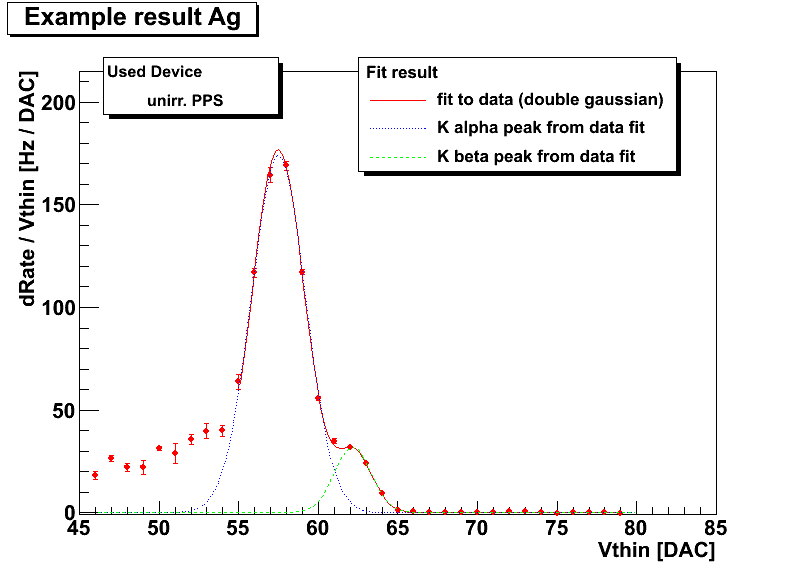}
        \label{fig:ag}
    }
    \label{fig:calibration}
    \caption{The pulser DAC calibration result (a) and the spectrum of Ag measured without using the analog hit information (b).}
\end{figure}

The pulser DAC calibration can be measured directly on the pads of the chip. The pulser DAC calibration result can be seen in figure \ref{fig:pulsercalib}.
The spectra of four x-ray sources were measured by measuring the hit rate as a function of threshold. The derivation of the resulting  "inverted" S-shaped curve is the spectrum. Figure \ref{fig:ag} shows an example of the measured Ag lines. A double gaussian fit was applied to the data points, the mean values of the fit were then correlated with the expected K$_\alpha$ and K$_\beta$ line of the spectrum. This calibration was also performed using Mo, Ba and Tb.

\subsection{Calibration constant measurement result}
\label{sub:calibmeasresult}

\begin{figure}[htp]
\centering
  \includegraphics[width=0.8\linewidth]{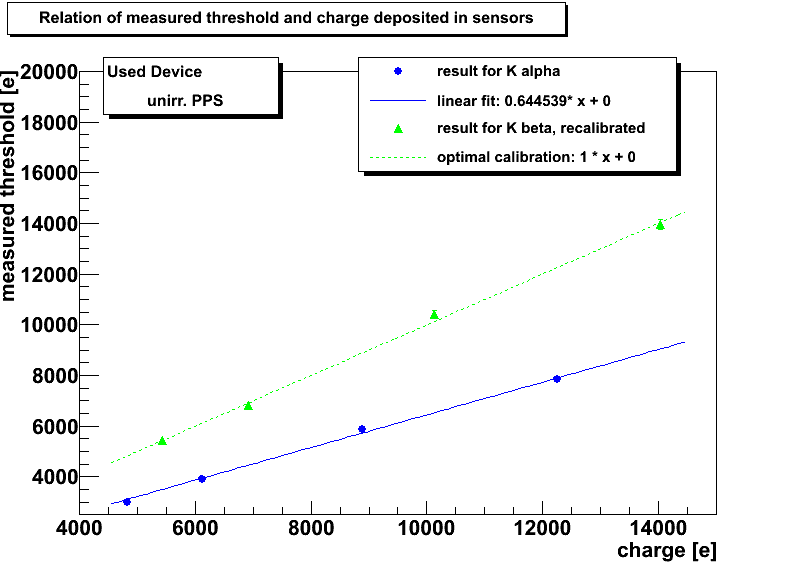}
  \caption{The measured peak energy in electrons using the design value for the injection capacitor in dependence of the charge generated in the sensor.}
  \label{fig:calibresult}
\end{figure}

The measured thresholds in the center of the $K_\alpha$ line peaks using the design value for the injection capacitor is plotted as a function of the charge generated by the photons in the sensor as blue dots in figure \ref{fig:calibresult}. A linear fit is applied and the injection capacitance can be calculated from the slope. The new calibration was then applied during the energy measurement of the $K_\beta$ peak and plotted in figure \ref{fig:calibresult} as green triangles. The dashed green linear function is not the result of a fit, but the linear function with slope equal to one and without any offset, because this is what gives to the correct calibration of the chip in this plot.
Obviously the data achieved with the new calibration nicely fits to the expectations. The injection capacitance when using both capacitors was measured to be 6.7 fF. The uncertainty of this measurement is expected to be in the order of 10 \%, and an independent measurement of the injection capacitance confirmed this result meanwhile. Please note that this kind of measurement was performed on a single sample only so far and it is still to be performed on a large amount of modules.

\section{Low threshold operation}
\label{sec:lowthresholds}

Low threshold operation is one of the key issues helping to achieve good hit detection efficiencies on highly irradiated sensors, when the amount of effective charge carriers seen by the preamplifier decreases. Lowering the threshold has the drawback of increasing the amount of fake hits (also called noise hits). A high amount of noise hits decreases the tracking performance of the detector. In the current \mbox{ATLAS} Pixel Detector all pixels with a noise hit probability per 25 ns (noise hit occupany, NOcc) higher than $10^{-7}$ are masked. Detailed studies investigating the minimum operational threshold with a good compromise between hit detection efficiency on irradiated sensors and NOcc per pixel using both module concepts are currently ongoing.

\subsection{Electronics noise and ToT studies in low threshold operation}
\label{sub:noisetot}

\begin{figure}[htp]
\centering
    \subfigure[]{
        \includegraphics[width=0.8\linewidth]{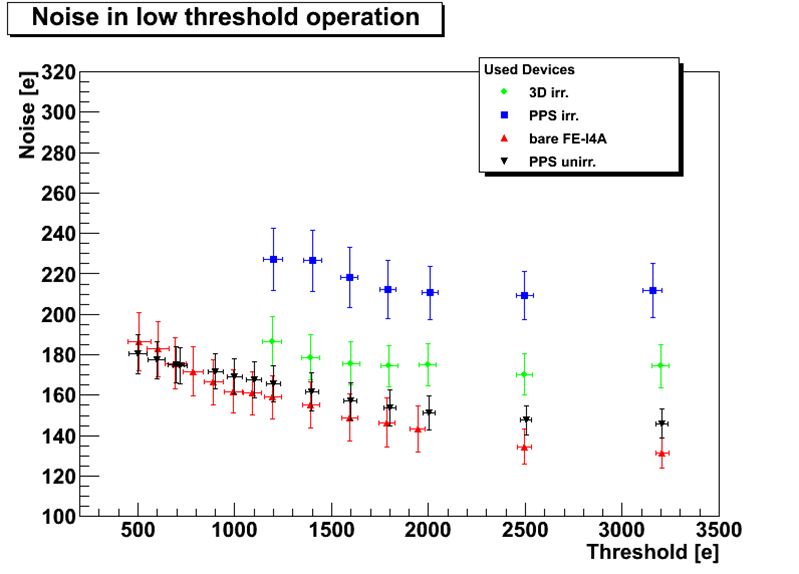}
        \label{fig:noise}
    }
    \subfigure[]{
        \includegraphics[width=0.8\linewidth]{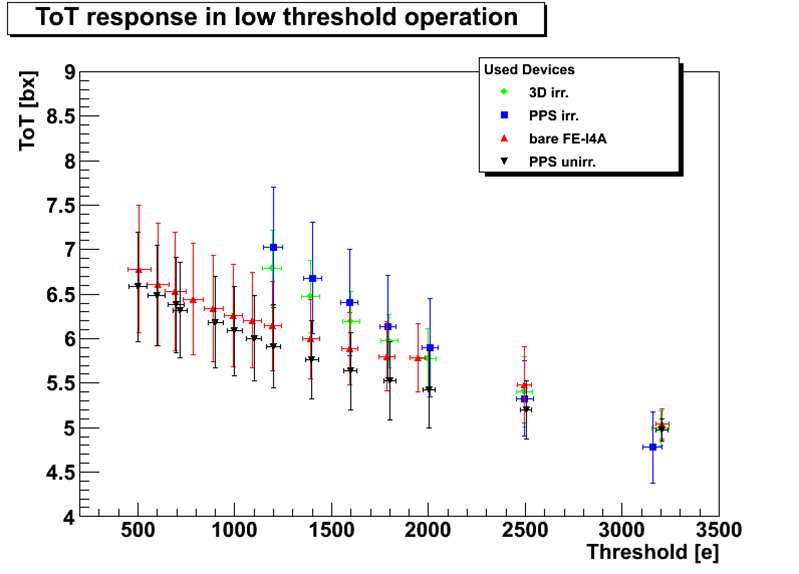}
        \label{fig:tot}
    }
    \label{fig:noisetot}
    \caption{The electronics noise (a) and analog hit information in terms of Time over Threshold in units of 25 ns (LHC bunch crossing clock) in low threshold operation (b).}
\end{figure}

The noise as well as the analog hit information in terms of time over discriminator threshold (ToT) in low threshold operation was studied. Figure \ref{fig:noise} shows a slightly increase of the measured electronics noise when going down with threshold. Note that due to inefficient cooling the irradiated planar device was operated with a very high leakage current of 3 mA explaining the high noise of this device. The increase for lower thresholds is observed in the unirradiated bare IC also, so the increase is expected to be a FE or measurement method effect.\\
The feedback current settings was not retuned during the measurement procedure in order to observe the expected increase of the ToT due to the triangular preamplifier pulse when lowering the discriminator threshold. This is shown in figure \ref{fig:tot}.

\subsection{Noise hit probability studies in low threshold operation}
\label{sub:nocc}

\begin{figure}[htp]
\centering
  \includegraphics[width=0.8\linewidth]{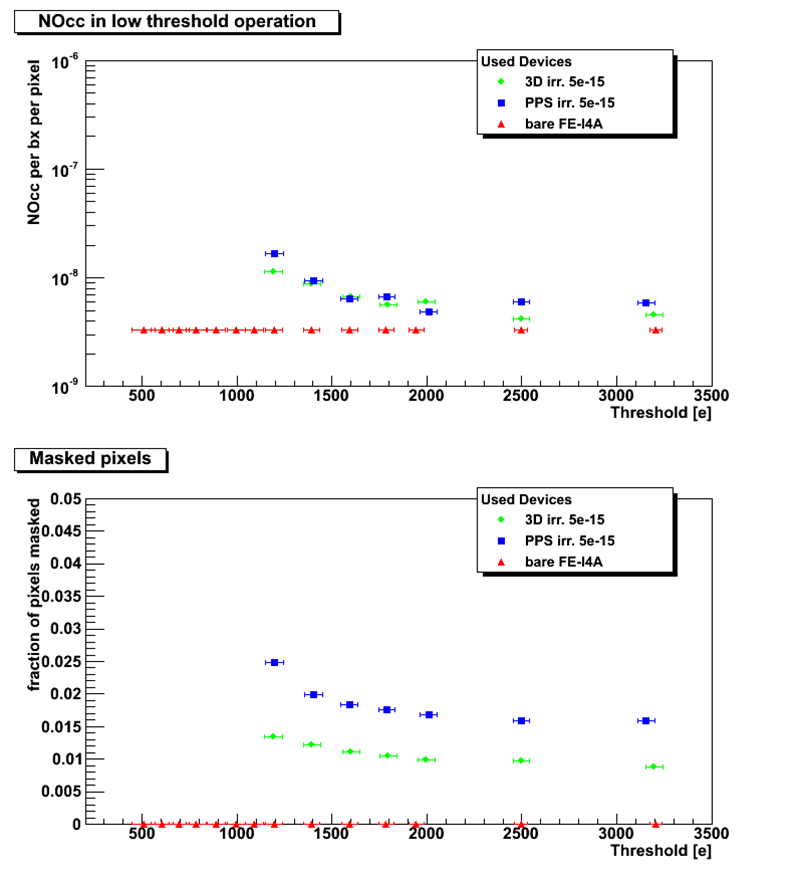}
  \caption{Noise hit probability per 25 ns and number of masked pixels as a function of threshold.}
  \label{fig:nocc}
\end{figure}

The top plot in figure \ref{fig:nocc} shows the threshold dependence of the NOcc for both module concepts after irradiation to $5 \cdot 10^{15}$ 1 MeV n$_{eq}$cm$^{-2}$ and for an unirradiated bare IC. The NOcc was measured sending $3 \cdot 10^8$ random triggers to the chip, which correlates to a sensitivity down to $3.34 \cdot 10^{-9}$ noise hits per pixel and 25 ns. All digitally problematic pixels have been masked offline, whereas all pixels showing an NOcc above $10^{-5}$ have been masked in analysis to avoid domination of single pixels with high NOcc in the calculation of the mean NOcc. The bottom plot in figure \ref{fig:nocc} shows the total amount of masked pixels in dependence of threshold. The 1-2 \% dead pixel fraction in the irradiated modules at high threshold is seen independently from the sensor type used. Preliminary investigations show this effect very likely to be caused by the very high dose in the front end electronics. To achieve the high fluence in the sensor using 26 MeV protons the chip has absorbed a total ionizing dose larger then 800 MRad, which is well above of the design irradiation tolerance of 300 MRad. Note that for modules irradiated using neutrons, the amount of dead pixels was not increased with respect to the performance before irradiation.\\
\begin{figure}[htp]
\centering
    \subfigure[]{
        \includegraphics[width=0.45\linewidth]{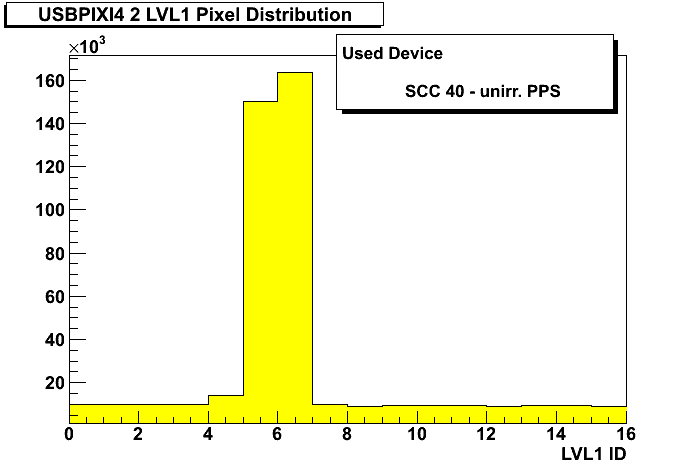}
        \label{fig:lv1before}
    }
    \subfigure[]{
        \includegraphics[width=0.45\linewidth]{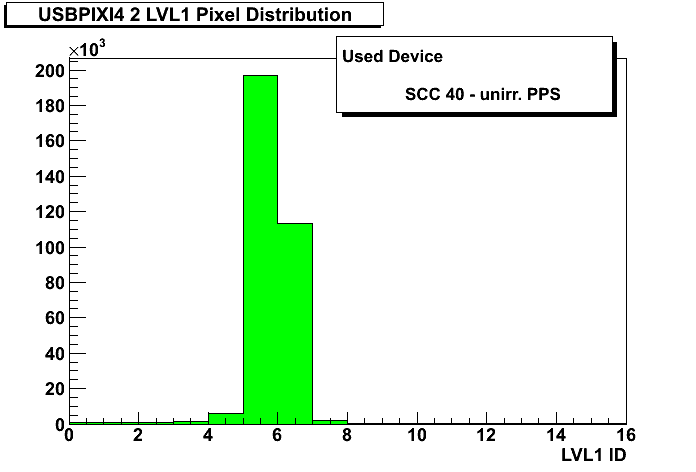}
        \label{fig:lv1after}
    }
    \label{fig:hittiming}
    \caption{Hit timing information (within a time window of 16 times 25 ns) before (a) and after (b) applying a noisy pixel mask. Obtained at \mbox{IBL} June test beam using 180 GeV pions.}
\end{figure}

 As the pixels showing a high NOcc do not change with time (fixed pattern noise), these measurements show that the chip can be operated independently of the sensor technology with a very low NOcc  of approximately $10^{-8}$ down to thresholds in the order of 1600 electrons without masking a large amount of pixels. Figure \ref{fig:lv1before} shows the hit timing information as obtained in the \mbox{IBL} June test beam at CERN without a noisy pixel mask applied. Clearly the large peak of hits recognized with fixed timing relation to the particles crossing the telescope is visible as well as a background of noise hits without fixed timing. A noisy pixel mask which disabled all pixels having a NOcc higher than $10^{-5}$ obtained using a procedure similar to the measurement used for the data in figure \ref{fig:nocc} was applied for data acquisition in figure \ref{fig:lv1after}. The background related to the noise hits disappeared, proving this procedure allows operation of the module with very low noise hit probability at low thresholds.

\section{Conclusions}
\label{sec:conclusions}
The new ATLAS pixel readout chip, FE-I4A has been characterized in great details using bare chips and bump bonded devices. All results show very promising performance regardless of the used sensor technology, planar n-in-n or double sided 3D silicon. Especially the low threshold operation of highly irradiated FE-I4A prototype modules with both sensor technologies is encouraging. Minor issues found in the bare IC characterizations have been changed in the design of FE-I4B and a charge injection calibration circuit has been added. A method to calibrate the FE-I4A injection circuit has been developed and will be used to test the new circuitry in FE-I4B. The FE-I4B will be the production chip for the ATLAS IBL and was submitted in September 2011. The planar pixel and 3D silicon sensor productions are ongoing. First IBL production modules are expected for beginning of 2012.


\end{document}